\documentclass[aip, jcp, reprint]{revtex4-1}

\usepackage{pgffor}
\usepackage{comment}
\usepackage{amsmath}
\usepackage{amsfonts}
\usepackage{graphicx}
\usepackage{comment}
\usepackage{xcolor}
\usepackage{epstopdf}
\usepackage{newtxtext,newtxmath}
\usepackage{physics}
\usepackage{hyperref} 
\usepackage{epstopdf}
\usepackage[T1]{fontenc}

\usepackage{soul}

 \hypersetup{
	colorlinks=true,       
	linkcolor=blue,          
	citecolor=blue,        
	filecolor=magenta,      
	urlcolor=blue           
}

\draft 

\begin{document}

\title{Many-body theory calculations of positronic-bonded molecular dianions}

\author{J. P. Cassidy}
\email{jcassidy18@qub.ac.uk}
\affiliation{Centre for Light-Matter Interactions, School of Mathematics and Physics, Queen's University Belfast,  Belfast BT7 1NN, Northern Ireland, United Kingdom}

\author{J. Hofierka}
\affiliation{Centre for Light-Matter Interactions, School of Mathematics and Physics, Queen's University Belfast,  Belfast BT7 1NN, Northern Ireland, United Kingdom}

\author{B. Cunningham}
\affiliation{Centre for Light-Matter Interactions, School of Mathematics and Physics, Queen's University Belfast,  Belfast BT7 1NN, Northern Ireland, United Kingdom}

\author{D. G. Green}
\email{d.green@qub.ac.uk}
\affiliation{Centre for Light-Matter Interactions, School of Mathematics and Physics, Queen's University Belfast,  Belfast BT7 1NN, Northern Ireland, United Kingdom}

\date{\today}

\begin{abstract}
The energetic stability of positron di-anion systems [A$^-;e^+;$A$^-$] is studied via many-body theory, where 
$A^-$ includes H$^{-}$, F$^{-}$, Cl$^{-}$ and the molecular anions (CN)$^{-}$ and  (NCO)$^{-}$. Specifically, the energy of the system as a function of ionic separation is determined by solving the Dyson equation for the positron in the field of the two anions, using a positron-anion self energy as constructed in [J. Hofierka, B. Cunningham, C. M. Rawlins, C. H. Patterson and D. G. Green, \emph{Nature} {\bf 606} 688 (2022)] that accounts for correlations including polarization, screening, and virtual-positronium formation. Calculations are performed for a positron interacting with H$_{2}^{2-}$, F$_{2}^{2-}$, and Cl$_{2}^{2-}$, and are found to be in good agreement with previous theory. In particular, we confirm the presence of two minima in the potential energy of the [H$^-;e^+$;H$^-$] system with respect to ionic separation: one a positronically-bonded [H$^-;e^+$;H$^-$] local minimum at  ionic separations $r\sim3.4$~\AA\phantom{}, and a global minimum  at smaller ionic separations $r\lesssim1.6$~\AA\phantom{} that gives overall instability of the system with respect to dissociation into a H$_2$ molecule and a positronium negative ion, Ps$^-$. The first predictions are made for positronic bonding in dianions consisting of molecular anionic fragments, specifically for (CN)$_{2}^{2-}$, and (NCO)$_{2}^{2-}$. In all cases we find that the molecules formed by the creation of a positronic bond are stable relative to dissociation into A$^-$ and $e^+$A$^-$ (positron bound to a single anion), with bond energies on the order of 1~eV and bond lengths on the order of several \r angstroms.
\end{abstract}


\maketitle 

\section{Introduction}
\label{sec: intro}

In their seminal 2018 paper, Charry, Varella and Reyes predicted that introducing a positron to two otherwise repelling hydride anions H$^-$ would lead to stabilization of the molecular complex via a \emph{positronic covalent bond} \cite{Reyes2018}.
Their analysis of the potential energy curves (PEC) calculated via the configuration interaction (CI) method for the [H$^-$;$e^+$;H$^-$] system revealed a minimum in the total system energy below the threshold for dissociation into a free hydride anion H$^-$ and a positron bound to the other anion, $e^+$H$^-$, demonstrating atomic matter fragments bound by antimatter.

The potential stability of the [H$^-$;$e^+$;H$^-$] system was subsequently also investigated in 
so-called ``multi-component Quantum Theory of Atoms in Molecules'' calculations
by Goli and Shahbazian\cite{Shahbazian2019}, and Monte Carlo calculations by Ito \emph{et al.} \cite{Tachikawa_Hbond2020}, and  by Bressanini\cite{Bressanini2021_Hbond}, with Bressanini showing that a local energy minimum occurred for anionic separations greater than 1.7~\AA\phantom{}, but that  the system was ultimately unstable against dissociation into H$_2$ and the positronium negative ion Ps$^-$. 

Since the original publication in 2018, there have been numerous studies of anionic bonding via antimatter. Moncada \emph{et al.}\cite{Reyes2019_dihalides} showed that positronic bonds could be formed in homo- and heteronuclear dihalide systems (they considered fluoride F$^-$, chloride Cl$^-$ and bromide Br$^-$ anions), and Ito \emph{et al.}\cite{Tachikawa_Libond2023} showed that a positron could stabilize two Li$^-$ anions.
Bressanini went on to show that two positrons could form a chemical bond in (PsH)$_2$\cite{Bressanini2021_TwoPosBond} in a similar way to how the electrons from two hydrogen atoms can form a bond to create the H$_2$ molecule, provided that the positrons involved have opposite spins (the conventional notation PsA is used to denote the positron-anion bound system \cite{SCHRADER1998}). However, analysis of the $^1\Sigma_g$ PEC for the (PsH)$_2$ molecule formed revealed that for internuclear distances less than 3.6~a.u. the molecule would dissociate into H$_2$ and Ps$^-$.
Charry \emph{et al.}\cite{Charry:2022} showed that a distinct non-electronic three-centre, two positron bond 
could be formed in a system containing two positrons and three hydride ions.

Recently, a number of us developed the many-body theory of positron binding to polyatomic molecules implemented via a Gaussian-basis representation in our {\tt EXCITON+} code \cite{Jaro22}. The approach enables the accurate account of the positron-molecule correlations including polarization, screening of the electron-positron Coulomb interaction, and the important process of virtual-positronium formation (where an electron temporarily tunnels to the positron). 
After successful benchmarking against explicitly correlated Gaussian and quantum Monte Carlo calculations for small molecules (LiH and formaldehyde respectively), it gave the first \emph{ab initio} calculations in agreement with over two decades of pioneering experiments, delineated the role of distinct correlations, and provided fundamental insight, e.g., the contribution of individual molecular orbitals and the role of the positron-molecule potenial anisotropy, and made predictions for systems including DNA bases. More recently it has provided \emph{ab initio} description of trends within molecular families\cite{Cassidy:2023}, quantified the effects of high-order diagrams to positron-atom scattering and annihilation properties \cite{JaroFrontiers2023}, and has been successfully extended to positron scattering and annihilation on non-binding molecules \cite{Charlie2023}. 

Here, we apply our many-body approach to study the formation of positronic bonds between selected anions. We first consider the H$_{2}^{2-}$, F$_{2}^{2-}$, Cl$_{2}^{2-}$ anionic systems, comparing with existing results. We then go beyond previous studies by making predictions of positronic bonding in dianion systems containing molecular anion fragments, specifically {(CN)$_{2}^{2-}$, and (NCO)$_{2}^{2-}$. 
} 
We find that all the systems are capable of forming positronic bonds that are stable relative to dissociation into A$^-$ and $e^+$A$^-$ (positron bound to a single anion), with corresponding bond energies on the order of 1~eV and bond lengths on the order of a few \r angstroms. In particular, we find that the positronically bonded [H$^-$;$e^+$;H$^-$] system is unstable relative to dissociation into H$_2$ and Ps$^-$ for small ionic separations $r\lesssim1.6$~\AA\phantom{}, agreeing with the previous work of Bressanini\cite{Bressanini2021_Hbond}. 

The paper is organized as follows. In Sec.~\ref{sec: theory} we present the theoretical details of the present study, including a brief overview of the many-body theory approach and numerical details for the present calculations. 
Section~\ref{sec: results} presents the results, including bond energy curves and the positron density at the bond length separation of each anion system, comparing to other theoretical studies where possible. We conclude in Sec~\ref{sec: sumary}.

\section{Theory}
\label{sec: theory}
Throughout this work we use atomic units unless otherwise stated, and employ the Born-Oppenheimer approximation for all systems under consideration. 

In this work we consider the stability of the [A$^{-};e^{+}$;A$^{-}$] systems relative to the dissociation [A$^{-};e^{+}$;A$^{-}]\to$ A$^{-}+\text{PsA}$, i.e. into one free anion and a positron bound to the other anion. As we discuss in Sec~\ref{sec: results}, at small ionic separations $r\lesssim1.7$~\AA\phantom{} the [H$^{-};e^{+}$;H$^{-}$] system's lowest energy dissociation channel is into H$_2+$ Ps$^-$\cite{Bressanini2021_Hbond}. At these small separations, the H$_2+$ Ps$^-$ PEC is below that of the [H$^{-};e^{+}$;H$^{-}$] PEC, resulting in the system not being absolutely stable. However, for $r>1.7$~\AA\phantom{} there is a local energy minimum of the [H$^{-};e^{+}$;H$^{-}$] system which is below the PECs of the other dissociation channels. We are able to compute this local minimum in this work using many-body theory, and we discuss it in order to benchmark against the results in the existing literature.

\begin{figure}[t!]
	\includegraphics[width=0.45\textwidth]{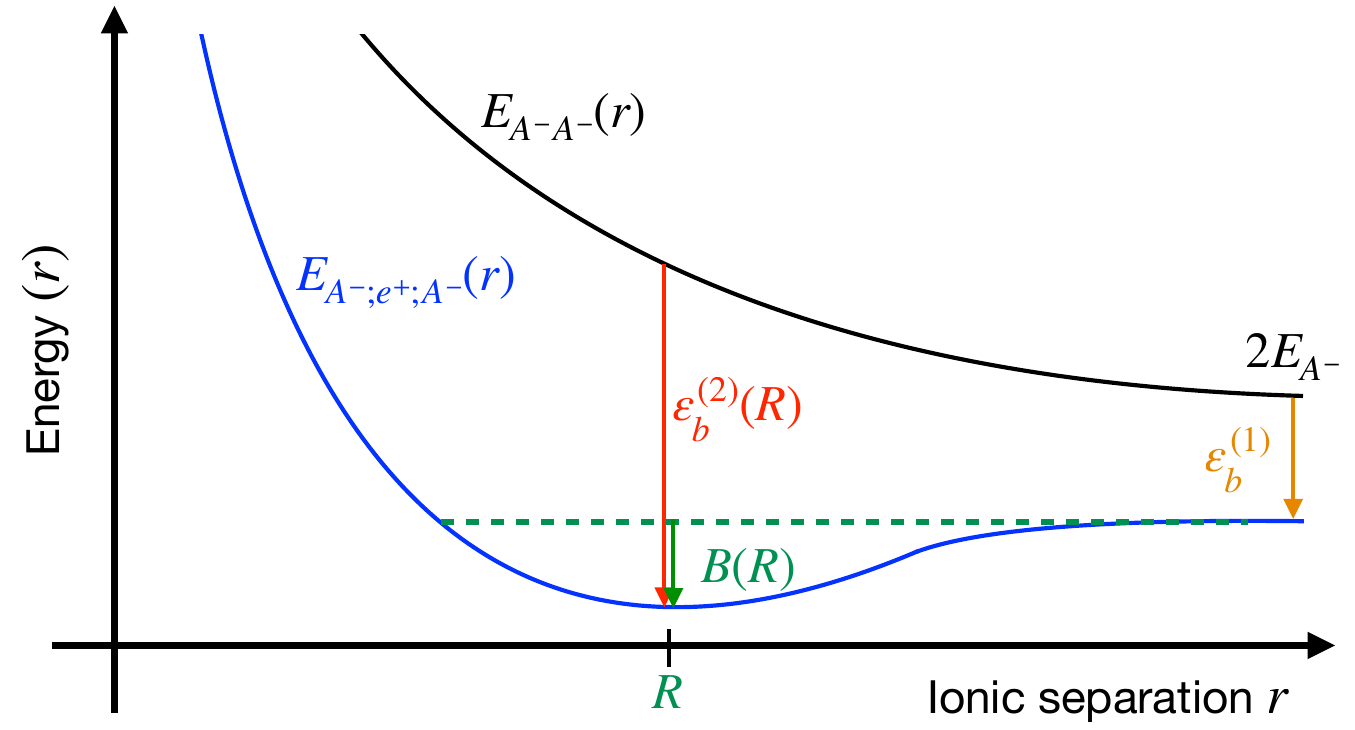}
	\caption{Schematic of the potential energy curves of positronic-bonded dianions, as in Eq.~(\ref{eq bond energy}). $E_{\text{A}^{-}\text{A}^-}(r)$ (black line) is the PEC of two anions separated by a distance $r$ in the absence of the positron, which in the limit of infinite $r$ has the threshold energy of twice the energy of a single anion $E_{\text{A}^-}$. The blue curve is for the positron-dianion system, which in the limit of infinite $r$ has the threshold energy of a free anion A$^-$ and PsA. $\varepsilon_b^{(2)}(r)$ is the positron binding energy to the dianionic molecule at separation $r$, while $\varepsilon_b^{(1)}$ is the positron binding energy to a single anion. $R$ is the bond length, i.e., the point at which $E_{\text{A}^{-}\text{A}^-}(r)$ is minimum.}
	\label{fig PECs}
\end{figure}

Assuming the lowest dissociation channel is to A$^-+$PsA, the  \emph{bond energy} of the positron and molecular dianion system is
\begin{eqnarray}\label{eq bond energy}
	B(r)&=& E_{\text{A}^-;e^+;\text{A}^-}(r) - E_{\text{PsA+A}^-},
\end{eqnarray} 
where  $E_{\text{PsA+A}^-}$ is the total energy of a free anion A$^-$ and PsA, and $E_{\text{A}^-;e^+;\text{A}^-}(r)$ is that of the positron-dianion system.
Thus, negative positronic bond energies indicate a stable complex relative to dissociation into the free anion A$^-$ and PsA, and $\lim_{r\to\infty}B(r)=0$. 
The positronic bond length $R$ is the value of $r$ at which $B(r)$ has a  minimum. 
In practice we determine the total energy of the positron-dianion system as (see Fig.~\ref{fig PECs})\footnote{In this approximation we account for positron-anion correlations, but assume the electron orbitals are unaffected by the presence of the positron. The good agreement we obtain with other methods suggests an \emph{a posteorori} justification for this.}  
\begin{eqnarray}
 E_{\text{A}^-;e^+;\text{A}^-}(r) = E_{\text{A}^{-}\text{A}^-}(r)-|\varepsilon_b^{(2)}(r)|, 
\end{eqnarray}
and the positron-anion plus anion threshold energy as
\begin{eqnarray}
 E_{\text{PsA+A}^-} = 2E_{\text{A}^{-}}-|\varepsilon_b^{(1)}|.
\end{eqnarray}
Here we calculate the all-electronic energy terms $E_{\text{A}^{-}\text{A}^-}(r)$ and $E_{\text{A}^{-}}$ in the Hartree-Fock approximation. The positron binding energies to a single anion $\varepsilon_b^{(1)}$ and to two anions (as a function of the interionic distance)  $\varepsilon_b^{(2)}(r)$ are obtained using many-body theory \cite{dzuba_mbt_noblegas,Cederbaum-elecpos,Gribakin:2004,DGG_posnobles,DGG_hlike,DGG:2015:core,Jaro22} accounting for important correlation effects, via solution of the Dyson equation for the positron in the field of the anion (or atom or molecule)
\cite{FetterWalecka,MBTexposed,DGG_posnobles,Jaro22}

\begin{equation}
	(\hat{H}_0+\hat{\Sigma}_\varepsilon)\psi_\varepsilon(\mathbf r)=\varepsilon\psi_\varepsilon(\mathbf r). \label{eq dyson eqn}
\end{equation}
\begin{figure}[t!]
	\includegraphics[scale=0.2]{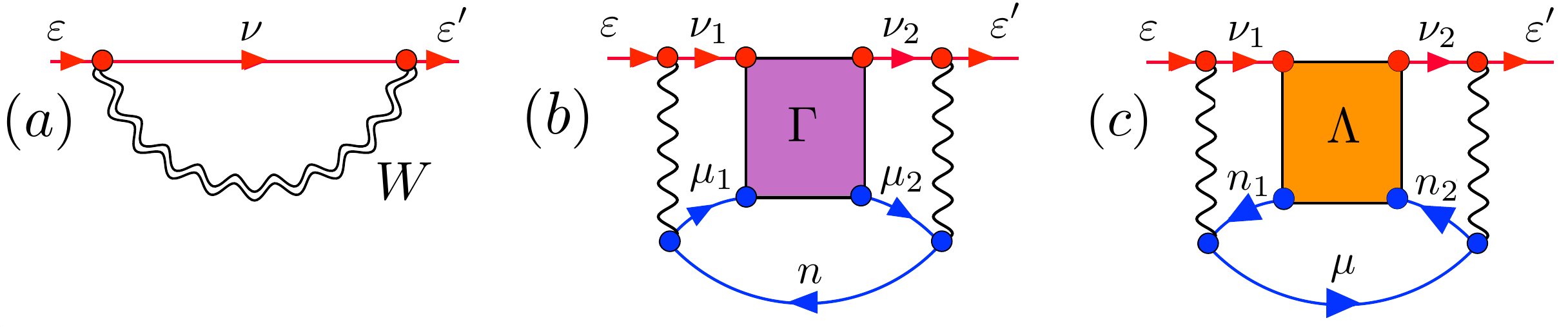}
	\caption{Diagrams the calculation of the positron self-energy. All lines directed to the right (left) represent Green's functions of particles (holes) propagating on the $N$ electron ground state of the anion(s). Single wavy lines represent Coulomb interactions. (a) The `$GW$' diagram, which is a product of the positron Green's function $G_{\nu}$ and the dressed Coulomb interaction $W$, accounts for polarization of the target by the positron, electron-hole interactions and screening of the Coulomb interaction; (b) the virtual-positronium formation diagram, which includes the infinite series of 
	electron-positron interactions (the $\Gamma$-block); (c) the positron-hole interaction diagram (the $\Lambda$-block). See [\citet{Jaro22}] for full details.} 
	\label{fig diagrams}
\end{figure}
Here, $\hat{H}_0$ is the zeroth order Hamiltonian of the positron in the static field of the anion(s), $\hat{\Sigma}_\varepsilon$ is the 
non-local, energy-dependent correlation potential that is equivalent to the irreducible self-energy of the positron in the field of the ion, $\psi_\varepsilon(\mathbf{r})$ is the positron (Dyson) wave function and $\varepsilon$ is the positron energy. 
For negative positron energy eigenvalues we define the positron binding energy as $\varepsilon_b=-\varepsilon$.  
We have calculated the positron binding energies $\varepsilon_b^{(1)}$ to H$^-$, F$^-$ and Cl$^-$ in \citep{Hofierka:atoms2023} and use those results here. 
Full details on the construction of $\Sigma$ and the solution of the Dyson equation, including numerical implementation, can be found in the recent work by some of the authors\cite{Jaro22}. 
Briefly, we expand the self energy $\Sigma$ diagrammatically in the residual electron-electron and electron-positron interactions.  We calculate three infinite classes of diagrams, shown in Fig.~\ref{fig diagrams}. The first is the $GW$ diagram, which we calculate in the Bethe-Salpeter Equation approximation, accounting for polarization of the electron cloud by the positron, screening of the Coulomb interaction (random phase approximation) and screened electron-hole interactions. 
The second is the virtual positronium (vPs) formation diagram that includes the infinite ladder series of electron-positron Coulomb interactions, the `$\Gamma$-block'. 
The third is the positron-hole repulsion diagram, which is of similar structure to the vPs diagram, but instead includes an infinite series of positron-hole Coulomb interactions.

We use the atomic aug-cc-pVQZ Gaussian basis sets of Dunning \cite{Dunning} to expand the electron and positron molecular orbital wave functions. These basis set centres coincide with the anionic centres. In order to ensure the convergence of the vPs contribution to the self energy, we employ additional hydrogen-type aug-cc-pVQZ bases on ``ghost atom'' centres to provide effective higher angular-momenta \cite{Swann:2018}.
For the [F$^{-}$;$e^{+}$;F$^{-}$], [Cl$^{-}$;$e^{+}$;Cl$^{-}$], [(CN)$^{-}$;$e^{+}$;(CN)$^{-}$], and [(NCO)$^{-}$;$e^{+}$;(NCO)$^{-}$], systems we used a single ghost centred between the anions along the bond axis. For the [H$^{-}$;$e^{+}$;H$^{-}$] system we used 5 ghosts: one centred between the two anions on the bond axis, and 4 more placed 1\AA\phantom{} away from the bond axis forming a square plane normal to the bond axis. We used the additional ghosts in the [H$^{-}$;$e^{+}$;H$^{-}$] system in order to compensate for the lack of $g$-type angular momentum functions in the H aug-cc-pVQZ basis which are present in the corresponding F, Cl, C, N, and O basis sets.

\section{Results and discussion}
\label{sec: results}
\begin{figure*}[ht!]
	\includegraphics[scale=0.3]{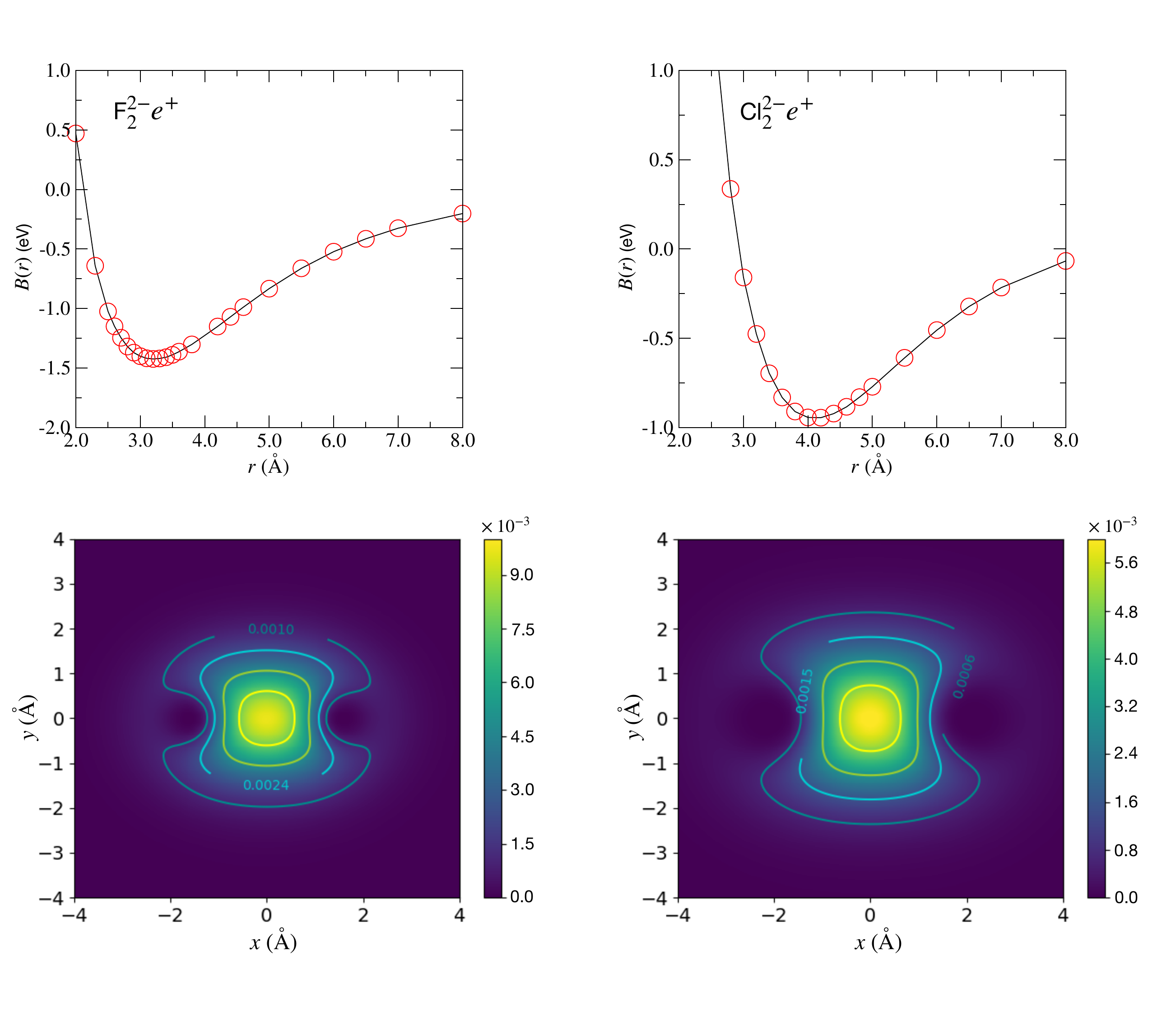}
	\caption{Positronic bond energy curves for the [F$^{-}$;$e^{+}$;F$^{-}$] and [Cl$^{-}$;$e^{+}$;Cl$^{-}$] systems, with negative bond energies indicating a stable complex with respect to dissociation into a free anion A$^-$ and PsA. The bond energy is defined in Eq.~(\ref{eq bond energy}). The position of the bond energy minimum defines the positronic bond length, $R$. Bottom panels show the positron Dyson wave function density (in a.u.), $|\psi_\varepsilon(\mathbf{r})|^2$ in the $xy$-plane, at the bond length separation $R$. Contours are drawn at 10\%, 25\%, 50\% and 80\% of the maximum density values for each system.      }
	\label{fig results}
\end{figure*}

The total energy curve for the [H$^{-}$;$e^{+}$;H$^{-}$] system, along with the positron wave function densities at the total energy minima are shown in Fig.~\ref{fig Hydrogen}. The bond energy curves and the positron wave function densities at the bond energy minima for the
 [F$^{-}$;$e^{+}$;F$^{-}$] and [Cl$^{-}$;$e^{+}$;Cl$^{-}$] systems are presented together in Fig.~\ref{fig results}. Those for [(CN)$^{-}$;$e^{+}$;(CN)$^{-}$], and [(NCO)$^{-}$;$e^{+}$;(NCO)$^{-}$] are shown in Fig.~\ref{fig CN NCO}, along with data for positron binding to the single anions (CN)$^{-}$ and (NCO)$^{-}$. We now discuss each system in turn.

\subsection{H$_{2}^{2-}$}
The positron plus hydrogen molecular dianion system has been studied in previous works by Charry \textit{et al.}\cite{Reyes2018}, Ito \textit{et al.}\cite{Tachikawa_Hbond2020} and Bressanini\cite{Bressanini2021_Hbond}. All confirmed the stability of the positronic-bonded [H$^-$; $e^+$; H$^-$] system with respect to dissociation into H$^-$ and PsH. 

\begin{figure*}[t!]
	\includegraphics[scale=0.49]{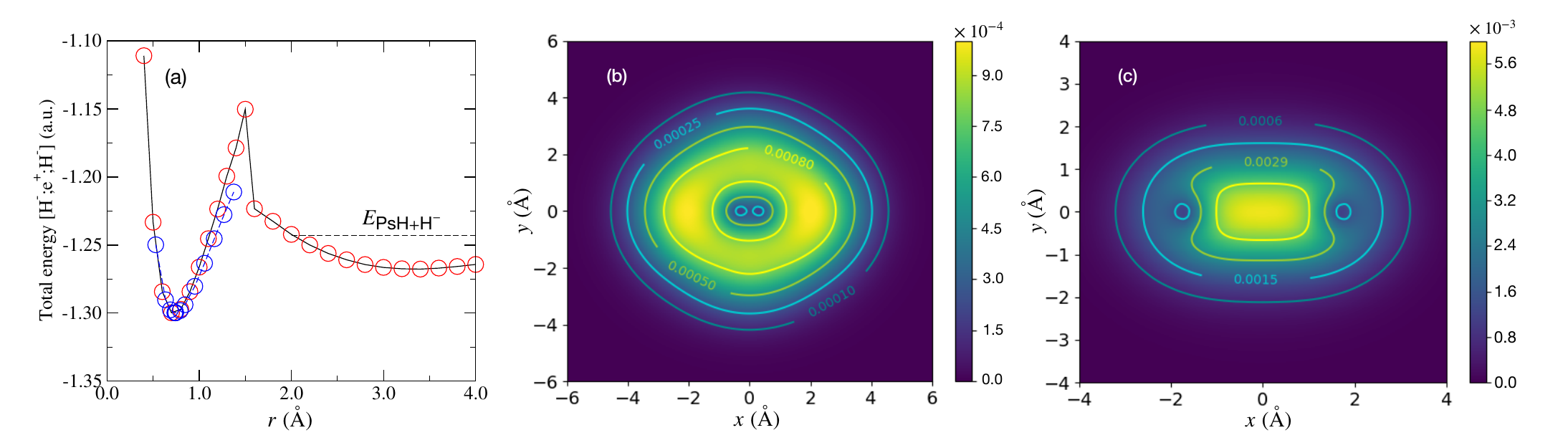}
	\caption{(a) The total energy of the [H$^-$;$e^+$;H$^-$] system as a function of the ionic separation distance $r$. The blue dashed curve is the PEC for the bare H$_2$ molecule, calculated by Kolos and Wolniewicz\cite{Kolos1968}, shifted vertically so that the minima coincide. The black dashed line is the PsH+H$^-$ energy threshold; (b) the positron Dyson wave function density (in a.u.) for H$_2^{2-}$, $|\psi_\varepsilon(\mathbf{r})|^2$ in the $xy$-plane, at the separation $r=0.7$~\AA\phantom{}. Contours are drawn at 10\%, 25\%, 50\% and 80\% of the maximum density value; (c) the positron Dyson wave function density (in a.u.) for H$_2^{2-}$, $|\psi_\varepsilon(\mathbf{r})|^2$ in the $xy$-plane, at the bond length separation $R=3.4$~\AA\phantom{}. Contours are drawn at 10\%, 25\%, 50\% and 80\% of the maximum density value.}
	\label{fig Hydrogen}
\end{figure*}

Figure~\ref{fig Hydrogen}a shows the total energy of the [H$^-$;$e^+$;H$^-$] system for ionic separations ranging from $r=0.4$~\AA\phantom{} to $r=4.0$~\AA.\phantom{} 
Two minima are evident, one at 0.7~\AA\phantom{} and one at 3.4~\AA\phantom{}, matching the so-called ``M1'' and ``M2'' minima described by Ito \textit{et al.}\cite{Tachikawa_Hbond2020} 

As shown in Fig.~\ref{fig Hydrogen}a and Table~\ref{tab:bonding}, at separations $r\gtrsim2$ \AA\phantom{} we find that the system is stable with respect to dissociation into H$^-$ and PsH, with a bond energy of magnitude 0.70~eV and equilibrium bond length $R=3.4$~\AA\phantom{}. The magnitude of the bond energy is in good agreement with the results of the aforementioned studies, e.g., the 0.68~eV full configuration interaction (FCI) and configuration interaction up to quadruple excitation (CISDTQ) results of Charry \textit{et al.}, the 0.61~eV result of the multicomponent diffusion  Monte Carlo (MC DMC) calculation by Ito \textit{et al.}, and the 0.64~eV diffusion Monte Carlo result of Bressanini. 
Our calculated bond length $R=3.4$~\AA\phantom{} is in very close agreement with the best results of Charry \textit{et al.}\cite{Reyes2018} and Ito \textit{et al.}\cite{Tachikawa_Hbond2020}, who both reported bond lengths of $R=3.26$~\AA,\phantom{} and in excellent agreement with Bressanini's range of $R=3.39\pm0.21$~\AA  \cite{Bressanini2021_Hbond}. The positronic density for the system at the $r=3.4$~\AA\phantom{} separation is shown in Fig.~\ref{fig Hydrogen}c, and shows a clear preference for the positron to be found in the intermediate region between the two H$^-$ anions. This is in perfect agreement with the original findings of Charry \textit{et al.}, that first presented such a positronic bond. 

\begin{table}
	\caption{\label{tab:bonding}
		Calculated positronic bond lengths $R$ (\AA) and the magnitudes of the bond energies $B(R)$ (eV) for positronic molecular dianions as in Eq.~(\ref{eq bond energy}). The bond energies are relative to dissociation into an infinitely separated free anion and a positron bound to the remaining anion [A$^{-}$;$e^{+}$;A$^{-}$]$\to$A$^{-}+\text{PsA}$.}
	\begin{ruledtabular}
		\begin{tabular}{llll}
			System & Method &  $R$  & $B(R)$      \\  \hline 
			H$_{2}^{2-}$
			   & FCI \cite{Reyes2018} & 3.26 & 0.68 \\
			   & CISDTQ \cite{Reyes2018} & 3.26 & 0.68 \\
			   & CBS \cite{Reyes2018} & 3.26 & 0.81 \\
			   & MC VMC \cite{Tachikawa_Hbond2020}   & 3.26 & 0.43 \\
			   & MC DMC  \cite{Tachikawa_Hbond2020}  & 3.26 & 0.61 \\
			   & VMC    \cite{Bressanini2021_Hbond}     & 3.39$\pm0.21$   & 0.62 \\
			   & DMC      \cite{Bressanini2021_Hbond}               &  3.39$\pm0.21$    & 0.64 \\
			   & \textbf{MBT (present work)} & \textbf{3.4} &\textbf{0.70 } \\
			   &&& \\
			   F$_2^{2-}$ 
			      & APMO/REN-PP3\cite{Reyes2019_dihalides} & 3.09 &1.13 \\
			      & \textbf{MBT (present work)} & \textbf{3.2} & \textbf{1.42} \\
			   
			   &&& \\
			   Cl$_2^{2-}$ 
			    & APMO/REN-PP3 \cite{Reyes2019_dihalides} & 3.87 & 0.86\\
			   & \textbf{MBT (present work)} & \textbf{4.2} & \textbf{0.94} \\
			      
			&&& \\
			   (CN)$_2^{2-}$ & \textbf{MBT (present work)} & \textbf{3.9} & \textbf{0.90} \\
			   &&& \\
			   (NCO)$_2^{2-}$ & \textbf{MBT (present work)} & \textbf{3.5} & \textbf{1.28} \\
		\end{tabular}
	\end{ruledtabular}
\end{table}

For separations $r<1.6$~\AA\phantom{} we found that positronium formation was no longer virtual, meaning that dissociation channels involving real Ps and Ps$^-$ were available. This corroborates the work of Bressanini \cite{Bressanini2021_Hbond}, who found that below 1.7~\AA\phantom{}, the lowest energy dissociation threshold involved formation of a Ps$^-$ anion, i.e., the positron was able to escape the system bound to electrons that originated from the bare anions. 
The sharp peak in the curve at $r\sim1.6$~\AA\phantom{} corresponds to the crossing of the H$_2$+Ps$^-$ and H$^-$+PsH PECs. 
Along with the total energy curve of [H$^-$;$e^+$;H$^-$], we also show the PEC of the H$_2$ molecule (originally calculated by Kolos and Wolniewicz\cite{Kolos1968}), shifted vertically so that the minimum of the H$_2$ PEC coincides with the [H$^-$;$e^+$;H$^-$] energy minimum. 
Given that the shapes of the curves are in very good agreement, this suggests that the [H$^-$;$e^+$;H$^-$] system has dissociated into bare H$_2$, with the two surplus electrons forming Ps$^-$ with the positron. 
We originally attempted to compare our [H$^-$;$e^+$;H$^-$] PEC with that of H$_2$ shifted down by the absolute value of the ground state energy of  Ps$^-$ (0.262~a.u.), and found that the resultant H$_2$+Ps$^-$ PEC had a minimum 0.137~a.u. below our [H$^-$;$e^+$;H$^-$] minimum. This discrepancy can be interpreted as the expectation energy of the kinetic energy of the Ps$^-$ being described (there will also be some discrepancy due to the fact that we only consider the all-electron energy terms in the Hartree-Fock approximation). 
Figure~\ref{fig Hydrogen}b shows the positron density at the $r=0.7$~\AA\phantom{} separation. In order to ensure an accurate description of the positronic density around the entire molecule, we used an extra two ghosts, each ghost 1~\AA\phantom{} away from the H$^-$ anions along the bond axis. In our description of the system at this separation, the positron is delocalized around the entirety of the molecule (but it is still bound to the system). The spatial distribution of the positron density around the entire molecule, coupled with the shape of the total [H$^-$;$e^+$;H$^-$] energy curve matching that of bare H$_2$,  provides further evidence that a Ps$^-$-like object has been formed.
We do note however that our description of such a Ps$^-$-like object is imperfect. Our $\Gamma$-block diagram contribution to the self-energy is only rigorous in cases where Ps formation is virtual, and strictly speaking it does not specifically allow for two electrons to tunnel near the positron. Furthermore, the use of Gaussian basis functions that decay quickly from the basis function centres mean we cannot describe the positron (and thus Ps$^-$) over all space far from the anions. However, a useful consequence of the decay of the Gaussian basis functions is the artificial confinement of the positron to regions of space near the anions, and this is why the many-body theory in its current implementation is able to at least \emph{partially} describe the Ps$^-$ system.

With regards to the experimental realization of the positronically-bonded [H$^-$;$e^+$;H$^-$] system, our results corroborate the suggestion by Bressanini\cite{Bressanini2021_Hbond} that the formation of the [H$^-$;$e^+$;H$^-$] complex could be evidenced by resonant scattering in collisions of Ps$^-$ with vibrationally excited H$_2$.

\subsection{F$_{2}^{2-}$}

We next consider halide anions. For these systems we are only aware of one other study by Moncada \textit{et al.}\cite{Reyes2019_dihalides} 
Owing to the more complicated many-electron structure of the halide anions and the resultant high computational cost compared to H$_{2}^{2-}$, their study used ``thermodynamic cycles'' to determine the positronic bonding energies in terms of positron binding energies to single anions and molecular anions, electron binding energies and the dissociation energies of purely electronic molecular dianions. Their positron binding energies were calculated using the APMO/REN-PP3 propagator method, which is a renormalized third-order approximation to the diagonal elements of the self energy. In contrast, our present many-body theory approach considers the full self energy of the positron, i.e., including off-diagonal elements in the matrix elements of  the diagrams in Fig.~\ref{fig diagrams}. 

Our bond energy curve for the F$_{2}^{2-}$ system is shown in Fig.~\ref{fig results}. Our results confirm that the system is stable with respect to dissociation into F$^-$ and PsF, with a bond energy of magnitude  1.42~eV and a bond length of 3.2~\AA.\phantom{} The bond energy is in reasonable  agreement with the result of Moncada \textit{et al.,} 1.13~eV, with our reported bond length in very good agreement with their reported 3.09~\AA.\phantom{} Figure~\ref{fig results} also shows the positronic density for the system, with the positron most likely to be found in the region between the F$^-$ anions, indicating a positronic bond.

\begin{figure*}[ht!!]
	\includegraphics[scale=0.265]{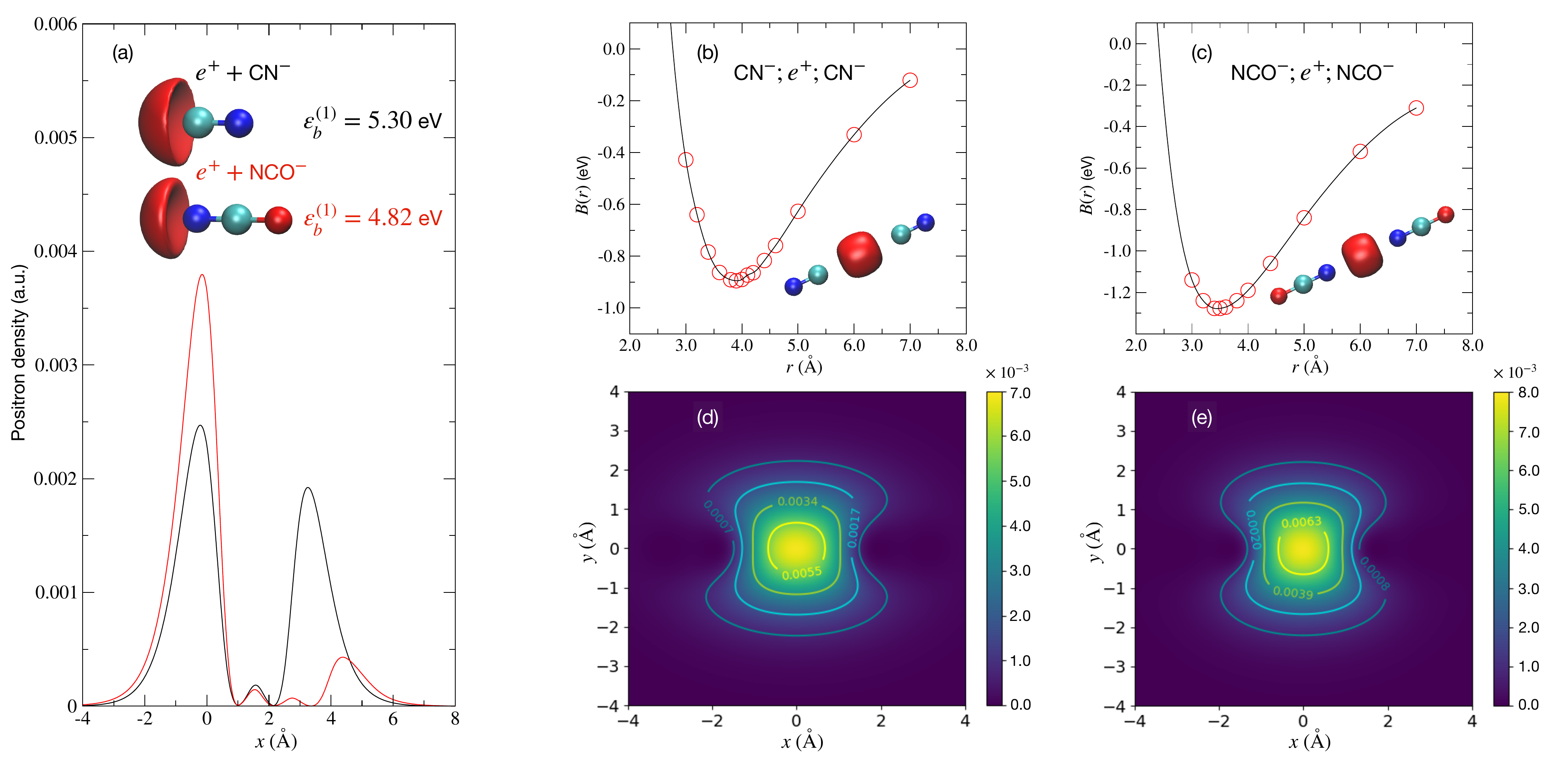}
	\caption{(a) The positron density along the molecular axis of a single CN$^-$ (black curve) and NCO$^-$ (red curve) molecular anion. Also shown are the positron binding energies for these molecular anions, $\varepsilon_b^{(1)}$, and the corresponding Dyson wave functions at 90\% of the wave function maximum. The locations of zero positron density coincide with the positions of the nuclei in each molecular anion (in the order as shown in the ball-and-stick models); (b) the positronic bond energy curve for the (CN)$_2^{2-}$ system along with the positron Dyson wave function at 80\% of the maximum wave function value. $r$ is taken to be the distance between the directly opposing C atoms; (c) the positronic bond energy curve for the (NCO)$_2^{2-}$ system along with the positron Dyson wave function at 80\% of the maximum wave function value. $r$ is taken to be the distance between the directly opposing N atoms; (d) the positron Dyson wave function density (in a.u.) for (CN)$_2^{2-}$ , $|\psi_\varepsilon(\mathbf{r})|^2$ in the $xy$-plane, at the bond length separation $R=3.9$~\AA\phantom{}. Contours are drawn at 10\%, 25\%, 50\% and 80\% of the maximum density value; (e) the positron Dyson wave function density (in a.u.) for (NCO)$_2^{2-}$ , $|\psi_\varepsilon(\mathbf{r})|^2$ in the $xy$-plane, at the bond length separation $R=3.5$~\AA\phantom{}. Contours are drawn at 10\%, 25\%, 50\% and 80\% of the maximum density value.       }
	\label{fig CN NCO}
\end{figure*}

\subsection{Cl$_{2}^{2-}$}
Our final comparison with a previously studied system is Cl$_{2}^{2-}$, which also featured in the study of Moncada \textit{et al.}\cite{Reyes2019_dihalides} We again confirm the stability of the system with respect to dissociation into Cl$^-$ and PsCl, and report a positronic bond energy of 0.94~eV, in very close agreement with the 0.86~eV reported by Moncada \textit{et al.} There is also good agreement in the reported bond lengths, with the present MBT bond length of 4.2~\AA\phantom{} less than $10$\% larger than the 3.87~\AA\phantom{} reported by Moncada \textit{et al.} It is important to note that both studies agree in that the positronic F$_{2}^{2-}$ system is more energetically stable than the positronic Cl$_{2}^{2-}$ system, demonstrating consistency in the predictions made by both methods. As shown in Fig.~\ref{fig results}, the positron is again most likely to be found in the region between the chloride anions, confirming the formation of a positronic bond.

\subsection{(CN)$_{2}^{2-}$ and (NCO)$_{2}^{2-}$}
We now turn our attention to  the systems comprised of CN$^{-}$ and NCO$^{-}$ anions. To our knowledge, our calculations for these systems are not only the first for these systems in the context of positronic bonding, but they are also the first calculations of positronic bonding where the bonded anionic fragments are themselves molecular in nature. We found (via many-body calculations using the full self energy of Fig.~\ref{fig diagrams}) that a single CN$^{-}$ anion binds the positron with a binding energy of $\varepsilon_b^{(1)}=5.30$~eV, with the maximum positron density localised near carbon.
For a single NCO$^{-}$ anion we calculated the positron binding energy to be $\varepsilon_b^{(1)}= 4.82$~eV, with the maximum positron density localised near nitrogen. The positronic density along the molecular axis for both anions is shown in Fig.~\ref{fig CN NCO}a.

Considering now the molecular dianions, we orientate the anionic fragments such that the molecular axes are collinear, with the carbon atoms of CN$^-$ facing each other for [(CN$^-$);$e^+$;(CN)$^-$], and the nitrogen atoms of NCO$^-$ facing each other for [(NCO$^-$);$e^+$;(NCO)$^-$] (see Figs.~\ref{fig CN NCO}b-c). For these systems the separation distance $r$ in the figures is the distance between the directly facing C or N atoms. We confirm the stability of the systems with respect to dissociation as shown in Figs.~\ref{fig CN NCO}b-c. For (CN)$_{2}^{2-}$ we report a bond energy of 0.90~eV and corresponding bond length of 3.9~\AA.~ For (NCO)$_{2}^{2-}$ we report a higher bond energy of 1.28~eV and a corresponding bond length of 3.5~\AA.~ The positron density in the $xy$ plane for (CN)$_2^{2-}$ is shown in Fig~\ref{fig CN NCO}d, and for (NCO)$_{2}^{2-}$ in Fig.~\ref{fig CN NCO}e. Both plots illustrate the positronic bond, consisting of positron density between the anionic fragments.
In the cases of (CN)$_{2}^{2-}$, and (NCO)$_{2}^{2-}$, there still remains the question of whether the positronically-bonded systems are stable with respect to dissociation to Ps and Ps$^-$, which would be of interest for future works.

\section{summary and conclusions}
\label{sec: sumary}
Many-body theory was applied to study the phenomenon of positronic bonding in molecular dianions.
For the [H$^{-}$;$e^{+}$;H$^{-}$], [F$^{-}$;$e^{+}$;F$^{-}$], and [Cl$^{-}$;$e^{+}$;Cl$^{-}$] positron-dianion systems the calculated bond energies and bond lengths were found to be in good agreement ($\lesssim$10\% error) with previous theoretical studies. 
For [H$^{-}$;$e^{+}$;H$^{-}$] we also observed the system to be unstable to dissociation into H$_2$ and Ps$^{-}$, corroborating prevoius studies \cite{}. 
We presented predictions for systems of positronic  bonding of \emph{molecular} fragments [(CN)$^{-}$;$e^{+}$;(CN)$^{-}$] and [(NCO)$^{-}$;$e^{+}$;(NCO)$^{-}$], finding positron-bonded complexes stable relative to the anion A$^-$ and PsA dissociation threshold. For these calculations we arranged the molecular fragments such that they were collinear and the positron-attracting ends of the single anions were directly facing each other. This orientation was chosen based on the fact that in a single CN$^-$ anion the positron was localized mainly on the C atom while in a single NCO$^-$ anion the positron was localized mainly on the N atom. It is possible that this orientation is not the most energetically preferable for the dimer system, either with or without the positron. Therefore, it may be of interest to investigate the stability of both dianion systems as a function of, for example, two orientation parameters-- one being similar to the separation parameter, $r$, as discussed in this paper, and an angular parameter to describe rotations of one of the fragments away from the linear orientation used here.

The many-body theory calculations used Hartree-Fock dianion potential energy curves. Improved calculations could be performed by accounting for eletron-electron correlations, e.g., in a $GW$@BSE calculation of the total energies, or via a self-consistent calculation in which the electron orbitals are allowed to adapt to the presence of the positron.  
Further studies of interest (beyond the orientation dependence of the stability of the dimer systems as described above) could include considering larger dianion molecular complexes, the rovibrational dependence and electronic-vibrational coupling in the dianion molecular fragment systems, and time-dependent studies of ionization, electron and positron scattering, and nuclear dynamics on such positronic-bonded complexes. Beyond fundamental interest, understanding these interactions could elucidate positron physics and chemistry in e.g., the interstellar medium \cite{RevModPhys.83.1001}, where anions are known to be present \cite{Millar:2017}.

\acknowledgements{
We thank Gleb Gribakin for useful discussions. JPC is funded by a Department for the Economy Northern Ireland postgraduate research studentship. This work was funded by the European Research Council grant 804383 'ANTI-ATOM' (DGG).}


%

\end{document}